\documentclass[twocolumn]{jetc20132e}

\usepackage{amsmath,graphicx}

\title{ 
Thermodynamics of continua:\\
the challenge of universality
}

\author{Peter V\'an$^{1,2,3}$ 
    \affiliation{ $^1$Dept. of Theoretical Physics, Wigner Research Centre for Physics, Institute for Particle and Nuclear Physics, \\  
      H-1525 Budapest, Konkoly Thege Mikl\'os \'ut 29-33, Hungary; \\
    $^2$Dept. of Energy Engineering, Budapest Univ. of Technology and Economics,\\
      H-1111, Budapest, M\H uegyetem rkp. 3-9,  Hungary; \\
    $^3$Montavid Thermodynamic Research Group \\
Email: van.peter@wigner.mta.hu}} 

\date{
{\small 
The explanation of the apparent universality of thermodynamics points toward the extension of the usual conceptual background of the second law. Arguments are collected that a basic guiding idea of stability of thermodynamic equilibrium combined with a proper interpretation of the entropy principle may provide the necessary solid foundation with verifiable consequences. 
}}

\begin{document}

\maketitle

\section{Introduction}

When treating the conceptual background of the second law, it is reasonable to start from the foundations, analysing the principles behind the concepts. 

The basic mystery in thermodynamics is the \textit{universality}. The validity of thermodynamic equations and theories regularly exceed the expectations. There are three independent aspects here:

\begin{enumerate}
 \item \textit{Uniformity}. We expect uniform principles and clear transition methods between the modeling levels. The validity of the second law is accepted in
  \begin{itemize}
    \item a) Thermostatics, treating the relation of state variables, 
    \item b) Ordinary thermodynamics, when processes of homogeneous bodies are modeled by time dependent state variables,
    \item c) Continuum thermodynamics, where the thermodynamic quantities are fields,
  \end{itemize}

\item \textit{Overdisciplinarity.} The concept of entropy and temperature appears from black holes to quark-gluon plasma, from general relativity to quantum chromodynamics. 

\item \textit{Mechanism independence}. The validity of the second law is independent of the particular mechanisms behind. Statistical mechanics, kinetic theory can provide particular demonstrations, but no proofs for a general principle.
\end{enumerate} 
 
One may wonder and discuss how extensive the validity of these aspects is. The question is whether and how one can understand the origin of the observed overdisciplinarity considering the expected uniformity. We consider as a key aspect the mentioned attitude to the mechanism independence -- the generality. 

In the following, we outline more exactly the challenge and a possible program of validation. Our working hypothesis is that the second law is a general principle and this is the reason of the universality of thermodynamics. Therefore, we need a guiding general idea, a conceptual understanding and, at the same time, we need a working strategy to translate this understanding to proper mathematical formulation of physical theories.

\section{The second law is material stability}

A guiding general idea cannot and must not postulate the existence of entropy, neither the increase of entropy: the aim is to introduce the physical origin of the entropy concept. The general idea cannot introduce statistical concepts because that violates the assumption of mechanism independence. Fluctuations or periodic machines are too specific. A general idea must be transparent. The general idea should produce a benchmark, a method of verification. A good idea should have a way of exact formulation in addition to flexibility. My suggestion is that 
\vskip 0.2cm
\textit{
\centerline{Thermodynamic equilibrium of simple materials}  
\centerline{is stable under neutral conditions.}}
\vskip 0.2cm

The idea that stability is connected to the second law, is ancient, it appears in the thermodynamic literature from different points of view and in different contexts (see, e.g. \cite{ColMiz67a,GlaPri71b,Gur75a,Daf79a,CasLeb82a,Ber86a}). Sometimes there are wrong connected claims. We do not state here the stability of steady-states (see, e.g. \cite{GlaEta74a,KeiFox74a} etc.). The validity of an idea can be discussed \cite{MulWei12a} and exact formulations of the statement are necessary. However, it is reasonable to believe that dissipation leads to stability of isolated simple materials. Without stability there is no observation, no reproduction of experiments. The above statement should be considered as a guiding idea, a challenge, a starting point of a program for the search of exact conditions \cite{PriStr95b}. This is not an exact statement yet, this is a principle.

\begin{figure}
\begin{center}
       \includegraphics[width=0.45\textwidth]{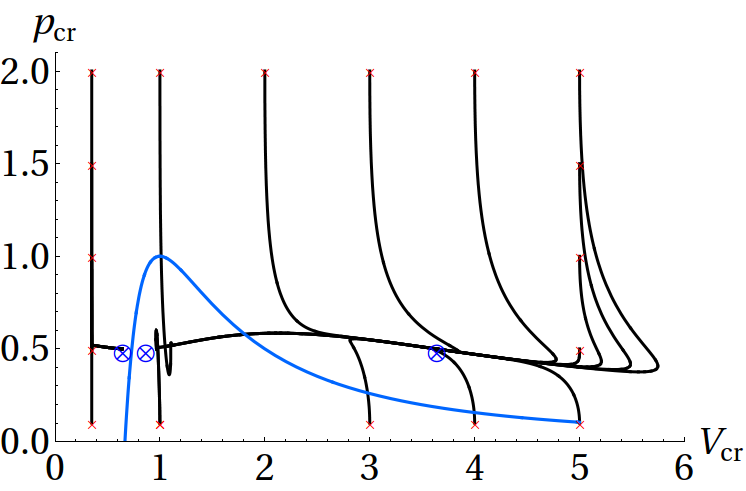}
\end{center}
\caption{ \label{vdWfig}
Irreversible processes in a homogeneous van der Waals gas in a cylinder closed by a piston are shown on the pressure-volume plane. The initial conditions form a rectangular area around the critical point. The three equilibrium points are denoted by crossed circles, the one under the spinodal is instable. The tendency toward the equilibrium indicates a slow manifold.}
\end{figure}

On the other hand, the stability concept of the second law is fully compatible to the other formulations. The complete thermostatics can be understood from this point of view. That entropy function is a potential in the thermodynamic phase space of classical homogeneous gases and fluids, that it is concave, and the requirement that entropy increases along several reasonable sets of differential equations, form the three conditions of a Lyapunov theorem. The thermodynamic equilibrium  of simple ordinary thermodynamic systems is asymptotically stable. This is a simple but rigorous result, developed thoroughly in the book of Matolcsi \cite{Mat05b}. There are clear conditions when it is valid and this is the necessary step between thermostatics and continuum thermodynamics claimed e.g. in \cite{TruBha77b}. Instead of further details, I show here some integral curves of a van der Waals body in a constant environment at the pressure-volume plane with critical state normalization.  These are processes, initial 
conditions are around the 
critical point indicated by crosses. The van der Waals gas body has a fold bifurcation at the critical point, and we can observe a slow manifold in Fig. \ref{vdWfig} with the particular interaction parameters.

In case of continua, our basic expectation is similar. Dissipation has to ensure that a homogeneous equilibrium is asymptotically stable in the absence of excitations and in case of neutral boundary conditions. Otherwise the dissipative theory is not properly constructed, it is a wrong model for real materials. Construction and validation are not really separable when speaking about principles. One can build a theory by any methods, introducing the empirical experience and also exploiting the entropy inequality and then check the stability of the homogeneous equilibrium. Does the entropy principle ensures the stability? The Fourier-Navier-Stokes system is linearly stable \cite{Van09a}, but generalized continua is not necessarily. Why? Is that a problem of the principle or of the formulation? 

Now we have arrived at the subject of the next section. To formulate a reasonable attitude, we should clarify the relation between the expected universality and the entropy principle in continuum physics.

\section{Entropy and universality}

A possible and rather usual understanding of the relation of thermodynamics and statistical theories assumes parallel  modeling levels according to the uniformity aspect of universality:
\renewcommand{\arraystretch}{1.5}
\begin{center}
\resizebox{\linewidth}{!}{
\begin{tabular}{c|c}
    Thermodynamics	    &  Statistical mechanics   \\\hline\hline
    Thermostatics	    & Equilibrium statistical mechanics  \\ \hline
    Ordinary thermodynamics  & Stochastic theory  \\ \hline
    Continuum thermodynamics & Kinetic theory \end{tabular}}\\
\vskip .21cm
{Table 1. Conceptual relation of thermodynamics and statistical physics}
\end{center}

There is also a relativistic version of these theories. There is a well-developed relativistic kinetic theory and also there are relativistic theories of fluids (some of them are unstable). A relativistic stochastic theory is a relatively new development \cite{DunHan09a}. However, our particular interest now is, that there is a \textit{relativistic thermostatics}, too \cite{Pla08a,Ein07a}. It is the statics of fast motion. It is an interesting subject in itself, but for us only one aspect is remarkable. One of the basic relations, the most widely accepted one, that leads to relativistic equilibrium statistical mechanics, e.g. to J\"uttner distribution, is the following:
$$
dE = TdS - pdV + vdG.
$$
This is a relation of energy $E$, temperature $T$, entropy $S$, pressure $p$ and volume $V$ of a thermodynamic body and its velocity $v$ and momentum $G$. This relativistic generalization of the Gibbs relation expresses the fact that energy and momentum cannot be separated. The content of the above formula is that entropy is a function of the volume and also of the energy-momentum four-vector $S=S(E^a,V)$. Only for an observer can entropy be a function of energy \textit{and} momentum separately. Relativistic thermostatics requires that entropy depends on momentum \cite{BirVan10a,VanBir12m}.

Quantum versions of statistical theories at the second column of Table 1 are well known. We have been struggling with quantum thermostatics for a long time, mostly via statistical approaches, too. However, what is \textit{quantum continuum thermodynamics}? It is originally not an outcome of a statistical theory. It is well-known for a long time that a special Korteweg fluid, where the pressure tensor is a function of the density gradient, the so-called Schr\"odinger-Madelung fluid, is equivalent to the one component Schr\"odinger equation \cite{Mad26a}. The fluid equations are the following: 
\begin{eqnarray}
\dot \rho + \rho\partial_iv^i &=& 0 \nonumber\\
\rho\dot v^i -\partial_j\left[\frac{\hbar^2}{8m^2}\left(\partial^{k}_{\, k}\rho\delta^{ij}+
\partial^{ij}\rho-\frac{2\partial^i\rho\partial^j\rho}{\rho}\right)\right] &=& 0,
\end{eqnarray}
where $\rho$ is the probability density, $m$ is the mass of the particle, $\hbar$ is the Planck constant and $v^i$ is the velocity field. The connection with the wave function is given by the Madelung transformation $\psi = \sqrt{\rho} e^{i I }$, where $I$ is related to the velocity potential in the simplest case $v_i =\frac{\hbar}{m}\partial_i I $. There are no complex fields, nor operators. Many researches extend the original analogy and put the Madelung idea into a wider context, and, at the same time, are speculating on the interpretational consequences. Some of the most interesting ones are \cite{Boh51b,Tak57a,Blo66b,Jan73a,Son91a,Hol93b,FulKat98a,Car07b}.   

This is not just an inconvenient side effect that can be forgotten and put aside. Quantum field theories, let they be Abelian or not, can be reformulated as fluid theories in general \cite{BisAta02a,JacEta04a}. There are vortices there, too. There is a corresponding thermodynamic background that requires a density gradient dependent entropy density \cite{VanFul06a}. For quantum continuum thermodynamics, the entropy density depends on the gradient of velocity.

Finally, one of the most striking relativistic  ---not yet observed--- phenomena is the Unruh effect. An accelerating observer may observe a thermal electromagnetic radiation of an oscillating charge.  In a covariant framework a thermodynamic theory of Unruh effect may require an acceleration dependent entropy function. As acceleration is related to gravitation one may wonder the role of the second law here... \cite{Ver11a}.

Therefore, the manifest overdisciplinary aspects of thermodynamic concepts indicate a need of a profound generalization of our classical approaches. Velocity and acceleration dependent entropies are well justified by relativistic theories. The traditional nonrelativistic concept of objectivity, which forbids this dependence, is wrong \cite{MatVan06a}. Moreover, gradient dependent constitutive state spaces and entropies are required for the explanation of quantum-hydrodynamic relations...

There is a simple idea that unfolds the mystery of universal aspects of thermodynamics. A theory is as universal as general the built-in assumptions and the conditions are. In continuum physics, the entropy principle is interpreted as an inequality, constrained by all other relevant conditions of the corresponding theory. Objectivity, material symmetries, kinematic restrictions, and fundamental balances are among the constraints that should be considered by the exploitation. This is a general approach to the second law if we analyse and properly apply the constraints and the other fundamental aspects \cite{ColNol63a,Liu72a,Cim07a}.

What are these fundamental aspects that should be scrutinized? It is already mentioned that the choice of the state space, both the basic and the constitutive one, is definitely one of them, where the known classical restrictions can be questioned. Moreover, entropy is a four-vector, entropy density and entropy current are frame dependent separations both in relativistic and in nonrelativistic spacetimes. Therefore, entropy current is a constitutive quantity, too.

These questions are connected. Assuming a classical entropy current, as the quotient of the heat flux and the temperature, $j_s^i = q^i/T$, one can prove that gradients are excluded from the state space \cite{Gur65a,AchSha95a,VanPap10a}. One may wonder that the multiple methods and ideas in weakly nonlocal continuum theories appear to circumvent these restrictions. These are for example the square-gradient ideas \cite{BedAta03a,KjeBed08b,GlaBed08a}, GENERIC \cite{GrmOtt97a,OttGrm97a,Ott05b}, phase fields theories \cite{HohHal77a,PenFif90a,AndEta98a}, different modifications of power \cite{Mau79a,DunSer85a,Mar02a,FabEta11a}, and others \cite{Cap85a,Gur96a}. Which is the best idea? Some researches say that the triumph of Copenhagen interpretation is due to pure manpower and the beautiful mathematical framework of von Neumann \cite{Cus00a} . What will happen in thermodynamics? Shall we able to discuss and reconcile  the problematic aspects? Or, at least, shall we able to understand each other?

My opinion is that universality is the key for ordering the different approaches and understanding their relations. Our best tool toward universal thermodynamic theories is the formulation of the entropy inequality as generally as it is possible and applying proper formulation of the additional principles, first of all objectivity, that determine the choice of the basic and constitutive state spaces. Therefore, if our methods of the second law are general and correct then we will obtain a universal theory.  

What is the role of stability then?

\section{Entropy and stability}

Thermodynamic equilibrium and thermodynamic state are delicate concepts. State variables should distinguish between the thermodynamic bodies, characterize the state and not the interaction \cite{BirVan11a}. The stability concept of the second law alone does not clarify the state variables and it is not constructive without the entropy principle. On the other hand, the entropy principle without stability is a complicated formulation of the  second law with an obscure physical content. Stability sometimes follows from the thermodynamic framework, but not always. Thermodynamic frameworks, the different entropy principles are not equivalent. 

The relation between stability and entropy is not simple. This is a long-discussed, deeply investigated and frequently rejected relation in classical continuum mechanics. A non-negative entropy production with concave entropy density alone does not ensure the asymptotic stability of equilibrium. Higher grade fluids are unstable \cite{DunFos74a,Jos81a}. At the local equilibrium level,a famous counterexample is the Eckart theory of relativistic fluids \cite{Eck40a3}. It is the simplest relativistic generalization of the Fourier-Navier-Stokes equations, constructed by thermodynamic principles. However, the homogeneous equilibrium of an Eckart fluid is violently unstable, in spite of the thermodynamic framework and nonnegative entropy production  \cite{HisLin85a}. 

Technically, the Lyapunov method for partial differential equations is not easy. It is simpler and more straightforward to check the linear stability of the equilibrium. Linear stability should be the consequence of the expected more general stability requirement and therefore serves as a convenient necessary condition, a suitable benchmark in the theory development. 

The universal extension of the entropy principle is a promising program. Recent examples of dissipative relativistic fluids indicate that an extension of the entropy principle may restore the expected stability \cite{VanBir08a,Van08a,Van09a,VanBir12a}. In this particular case, momentum also has to be among the state variables and, most importantly, the momentum balance is a constraint of the entropy inequality \cite{Van08a,VanBir12m}. 

The thermodynamic framework and the stability of homogeneous equilibrium in case of neutral conditions are two sides of the same coin. Stability is not only a general idea behind the second law, but also a verification tool of thermodynamic theories.

\section{How universal?}

The first book of the Landau-Lifshitz series of theoretical physics is about analytical mechanics \cite{LanLif76b}. It starts with a mind provoking derivation of the Lagrangian of a free point mass by spacetime symmetries and Hamiltonian variational principles. This is an attempt to understand the origin of evolution equations in physics. However, the variational principles as tools are not universal. Dissipation cannot be incorporated easily. Heat conduction and also dissipative mechanical systems in general cannot be understood with the help of variational principles, even the best attempts are artificial and their validity is restricted \cite{Gya70b,Nyi91a2,MarGam91a,VanMus95a,VanNyi99a}.

On the other hand, the previously outlined entropy principle provides a possibility to construct and derive evolution equations both for the nondissipative and the dissipative cases \cite{VanAta08a}. The clear examples in this respect are the evolution equations for internal variables, where restrictions from the second law provide the best way of construction, recovering and including results from dissipation potentials or variational assumptions, without these additional hypotheses \cite{Van09a1}. Moreover, in some investigated cases, the nondissipative part of the resulted evolution equation has an Euler-Lagrange form, the thermodynamic potential is connected to a Lagrangian. This generality requires the extension of the second law incorporating gradients of dual internal variables in the constitutive state space. This extension, the method of dual internal variables introduces a general framework of dissipative mechanical phenomena \cite{BerEta10a2,BerEta11a1,EngBer12a,BerBer13a}.

Generalized mechanics provides an example where several independent methods were applied for generating evolution equations of internal variables of mechanical origin \cite{CosCos09b,Min64a,EriSuh64a}. Here the thermodynamic method of dual internal variables ---the previous extension of the entropy principle--- can generate the evolution equations \cite{BerEta11a,VanEta13m}. The dissipative part of the evolution equations promisingly stabilizes the thermodynamic equilibrium according to linear stability analysis \cite{Van13m}. 

Then one may ask: how universal are the thermodynamic principles? As general laws of nature, their validity incorporates mechanics, electrodynamics and every discipline of physics \cite{Ver01p,Ver04a}. With a proper formulation and understanding, we can grasp the very origin and connection between these seemingly separated fields. This is the final dream of universality.

\section{Universal summary}

There are some principles in physics that provide a driving force for the development. Those are sometimes subjective feelings, like the requirement of simplicity and harmony. Others express deeper and almost inevitable requirements of physical theories, like objectivity. In a sense it is convenient to say that some expected general principles are invalid \cite{Mul72a,Rug12a,CarEta04a,Evans_news}. That way we may separate ourselves for some inconveniences and close a direction of investigations. This reductionism is frequently fruitful in focusing and deepening our understanding. On the other hand, general principles are pharoses in the scientific landscape, shining lights that keep ourselves on the right way. 
Focusing on the nearby stormy waves cannot prevent us from the greater dangers of reefs, which can be detected only from a right perspective. From this point of view an ostensible violation of a general principle is not only an indication that its validity is limited, but also a challenge that something is not properly formulated. 
 
The experienced universality indicates the need for the following generalizations of the entropy principle in continuum theories:
 \begin{itemize}
  \item -- The entropy four-vector is a constitutive quantity, both in classical and relativistic theories. Therefore, the entropy current density is constitutive.
  \item --  Momentum should be incorporated in the basic state space when dealing with mechanics and thermodynamics.
  \item -- Space and time derivatives of the basic variables cannot be excluded from the constitutive state space.
  \item -- The momentum balance must be considered directly as a constraint in the entropy inequality, beyond the internal energy concept. 
 \end{itemize}
These generalizations, in particular the clarification of the role of energy and momentum,  require the proper formulation and use of objectivity.
 
The tool of the validation of the mentioned generalizations ---a benchmark--- is provided by the stability concept of the second law. In particular, the linear stability of the evolution equations is a necessary condition.
 
Understanding the extent and the origin of the manifest universality of thermodynamic principles is one of the greatest challenges of thermodynamics.

\section{Acknowledgement}   
The work was supported by the grants OTKA K81161, K104260 and and T\'ET 10-1-2011-0061/ZA-15-2009.


\end{document}